\documentclass[12pt]{article}
\advance\topmargin by -1.6cm

\newenvironment{eq}
{\[\begin{array}}{\end{array}\]{}}

\let\rvec=\vec        

 \def\({\Bigl(}
\def\){\Bigr)}
 \def\|{\Big|}
\def\then{\Rightarrow}
 \def\o{\circ}

\def\x{\times}

\def\ox{\otimes}

\def\pl{{~\oplus~}}

\def\PL{\displaystyle \bigoplus}

\def\mid{\big\bracevert}

\def\subnoteq{\subset}

\def\supnoteq{\supset}
\def\and{\wedge}

\def\rin{{\,\in\kern-.42em\in}}
 
 \def\diag{{\,{\rm diag}\,}}

\def\spec{\,{\rm spec}\,}

\def\det{\,{\rm det }\,}
\def\id{\,{\rm id}}

\def\INV{\,\hbox{INV}}

\def\ad{{\,{\rm ad}\,}}

\def\A{{\,{\rm A\kern-.55emA}}}
\def\B{{\,{\rm I\kern-.2emB}}}
\def\C{{\,{\rm I\kern-.55emC}}}
\def\E{{\,{\rm I\kern-.2emE}}}
\def\G{{\,{\rm I\kern-.55emG}}}
\def\H{{{\rm I\kern-.2emH}}}
\def\I{{\,{\rm I\kern-.2emI}}}
\def\K{{\,{\rm I\kern-.2emK}}}
\def\L{{\,{\rm I\kern-.2emL}}}
\def\M{{\,{\rm I\kern-.16emM}}}
\def\N{{\,{\rm I\kern-.16emN}}}
\def\Q{{\,{\rm I\kern-.5emQ}}}
\def\R{{{\rm I\kern-.2emR}}}
\def\S{{\,{\rm I\kern-.42emS}}}
\def\T{{\,{\rm I\kern-.37emT}}}
\def\UU{{\,{\rm I\kern-.51emU}}}
\def\Z{{\,{\rm Z\kern-.32emZ}}}

\def\p{\partial}

\def\al{\alpha}  \def\be{\beta} \def\ga{\gamma}
\def\de{\delta}  \def\ep{\epsilon}

   \def\om{\omega} 
\def\phi{\varphi}
 \def\Ga{\Gamma}  
    \def\La{\Lambda}

\def\vec#1{\underline{\bf vec}_{#1}}

\def\GL{{\bf GL}}
\def\SL{{\bf SL}}
\def\U{{\bf U}}

\def\SU{{\bf SU}}

\def\SO{{\bf SO}}

 \def\D{{\bl D}}
\def\AL{{\bf AL}}

\def\angle#1{\langle#1\rangle}

\def\rstate#1{|#1\rangle}
\def\lstate#1{\langle#1|}

\def\ro#1{{\rm #1}}
\def\bl#1{{\bf {#1}}}
\def\cl#1{{\cal #1}}

\def\ol#1{\overline{#1}}

\def\dprod#1#2{\langle#1,#2\rangle}
\def\sprod#1#2{\langle#1|#2\rangle}

\def\com#1#2{\lbrack#1,#2\rbrack}

\def\acom#1#2{\{#1,#2\}}

\def\map{\longrightarrow}
\def\inmap{\hookrightarrow}
\def\lrmap{\leftrightarrow}

\def\mape{\longmapsto}

\def\dmape{\vcenter{\hrule\hbox{$\Big\downarrow$}}}

\def\Diagre#1#2#3#4{\matrix{ \noalign{\vskip0mm}
                                             &&\cr
         #1      &\mape       &   #2            \cr
     \dmape      &            &  \dmape         \cr
         #4      &\mape       &   #3            \cr
\noalign{\vskip5mm}             }}

\begin{document}
\begin{titlepage}
\hfill MPI-PhT/01-54


\centerline{\bf DEFINITE AND INDEFINITE UNITARY}
\vskip3mm
\centerline{\bf TIME  REPRESENTATIONS}
\vskip3mm
\centerline{\bf FOR HILBERT AND NON-HILBERT SPACES\footnote{\scriptsize
Talk presented on the `Worshop on Resonances and Time
Asymmetric Quantum Theory, Jaca(Spain), 30 May to 4 June (2001)}
}
\vskip15mm
\centerline{Heinrich Saller\footnote{\scriptsize
hns@mppmu.mpg.de}}
\centerline{Max-Planck-Institut f\"ur Physik and Astrophysik}
\centerline{Werner-Heisenberg-Institut f\"ur Physik}
\centerline{M\"unchen}
\vskip25mm

\centerline{\bf Abstract}
\vskip2mm
Stable states (particles), ghosts and unstable states (particles) are
discussed with respect to the time representations involved, their
unitary groups and the
induced Hilbert spaces.
Unstable particles with their decay channels
are treated as higher dimensional probability collectives with
nonabelian probability groups $\U(n)$
generalizing the individual abelian $\U(1)$-normalization
for stable states (particles).

\vfill

\vskip5mm

{\small \tableofcontents}
\vskip5mm
\end{titlepage}

\newpage

\section{Dynamics, Unitarity and Time Reflection}

A physical dynamics as a complex representation of time defines itself
an inner product, sometimes a Hilbert space product,
wherein a quantum dynamics is experimentally interpreted
with transition and probability amplitudes etc.
In this section, the mathematical background structure\cite{ALG9} is given.

As model for time
the abelian totally ordered  real numbers are used
- either multiplicatively, called time group
\begin{eq}{l}
\D(1)=\{e^t\mid t\in \R\}
\end{eq}or additively, called time Lie algebra\footnote{\scriptsize
The Lie algebra for a Lie group $G$ will be denoted by
$\log G$.}
with the
time translations
\begin{eq}{l}
\R=\log\D(1)=\{t\}\cr
\end{eq}Obviously, as Lie groups $(\D(1),\cdot)$ and $(\R,+)$
as isomorphic.

A quantum dynamics is a  representation
of time as group  in the automorphisms
$\GL(V)$ and as Lie algebra in the   endomorphisms $\AL(V)$
of a complex vector space $V$
\begin{eq}{rll}
\hbox{group:}&D:\D(1)&\hskip-2mm\map \GL(V),e^t\mape D(t),\left\{\begin{array}{rl}
D(0)&=\id_V\cr
D(t+s)&=D(t)\o D(s)\end{array}\right.\cr
\hbox{Lie algebra:}&\cl D:\R&\hskip-2mm\map \AL(V),t\mape \cl D(t)=t\cl D(1)
\end{eq}The represented basis of the time translations is - up to $i$ -
the Hamiltonian
\begin{eq}{l}
\cl D(1)=i H
\end{eq}States (bound states, scattering states, particles) have to be eigenvectors
under time action.

The complex representation space has to come with
a  conjugation in order to represent the realness of the Lie structure
with concepts like hermitian (real)-antihermitian (imaginary) etc.
A conjugation of a  complex vector space $V$
is an antilinear isomorphism
to its dual space $V^T$  (vector space with the linear $V$-forms)
\begin{eq}{l}
V\stackrel *\lrmap V^T,~~v\lrmap v^* ,~~v^{**}=v\cr
\hbox{Dirac notation: }v=\rstate v,~~v^*=\lstate v
\end{eq}For spaces with dimension $n\ge2$ there is not only the naive number
conjugation (canonical conjugation $\al\lrmap\ol\al$).
There may exist also more than one conjugation for a vector space, e.g.
the conjugation connecting creation and annihilation operators and the
conjugation connecting particles and antiparticles.

With a conjugation the $V$-endomorphisms $f\in \AL(V)$ can be conjugated too
by using the transposed endomorphisms $f^T:V^T\map V^T$
\begin{eq}{l}
\AL(V)\stackrel *\lrmap \AL(V)\hbox{ with }
[f:V\map V]\stackrel *\lrmap [f^*:V\stackrel *\map V^T
\stackrel{f^T}\map V^T \stackrel *\map V]\cr
\end{eq}The realness of the time group is implemented
by the hermiticity of the Hamiltonian -
with respect to the conjugation $*$
\begin{eq}{l}
H=H^*
\end{eq}Therewith, the generator $iH$
for the represented real time translations $\R$ is
antihermitian and the represented time
group unitary - always
with respect to the conjugation $*$
\begin{eq}{rll}
\cl D(1)^*&=\cl D(-1)=-\cl D(1)&\hbox{$*$-antihermitian for time translations}\cr
D(t)^*&=D(-t)=D(t)^{-1}&\hbox{$*$-unitary for group representation}\cr
\end{eq}The induced endomorphism  conjugation
implements the  time reflection
\begin{eq}{c}
\Diagre t{-t}{D(t)^*,\cl D(t)^*}{D(t),\cl D(t)}\cr
\end{eq}

With the dual product
(bilinear) for the vector space $V$ and its linear
forms
\begin{eq}{l}
V^T\x V\map\C,~~(\om,w)\mape\dprod\om w
\end{eq}a conjugation
is equivalent
to an inner product (a nonsingular sesquilinear form)
\begin{eq}{l}
V\x V\map\C,~(v,w)\mape\sprod v w=\dprod {v^*}w
\end{eq}

The  unitary group, characterizing
the conjugation $*$, is
 the invariance group of the induced inner product
\begin{eq}{l}
\U(V,*)=\{u\in\GL(V)\mid \sprod{u.v}{u.w}=\sprod vw\hbox{ for all }v,w\in V\}
\end{eq}This  defines the signature of the unitarity and of the conjugation
\begin{eq}{l}
V\cong\C^n,~~\sprod~~\cong{\scriptsize\pmatrix{\bl1_{n_+}&0\cr 0&-\bl1_{n_-}\cr}}
,~~n_++n_-=n,~~
\U(V,*)=\U(n_+,n_-)
\end{eq}There are as many different types of conjugations as there are
signatures -  for $n=1$ only $\U(1)$, for $n=2$ one has $\U(2)$ and
$\U(1,1)$ etc. With the exception of a Euclidean conjugation $\U(n)$, denoted with the five cornered star $\star$,
where one has a scalar product and a Hilbert space structure for the
vector space $V$,
all conjugations and associated
inner products are indefinite.
In the following time representations with both definite and indefinite
conjugations will be considered.

To become familiar with indefinite conjugations
a $\U(1,1)$ inner product is considered
\begin{eq}{l}
\hbox{for }\U(1,1):~~\sprod~~\cong
{\scriptsize\pmatrix{0&1\cr 1&0\cr}}
\sim {\scriptsize\pmatrix{1&0\cr 0&-1\cr}}
\end{eq}Us usual, the inner product matrix depends on
the vector space  basis: A diagonal matrix
with the explicit signature arises for Sylvester bases
whereas  neutral, not orthogonal
pairs with trivial norm show up in Witt bases. In the example above Sylvester
and Witt bases are related to each other with the automorphisms
$w={1\over\sqrt2}{\scriptsize\pmatrix{1&-1\cr 1&1\cr}}$.
The inner product defines the $\U(1,1)$-conjugation, denoted with $\x$,
which - in the basis with the skew-diagonal matrix for the
inner product -
interchanges the conjugated diagonal elements
in contrast to the familiar $\U(2)$-conjugation $\star$
(number conjugation and transposition)
which interchanges the
conjugated skew-diagonal elements
\begin{eq}{rll}
\U(2)=\star:&
{\scriptsize\pmatrix{\al&\be\cr\ga&\de\cr}}^\star
&={\scriptsize\pmatrix{\ol\al&\ol\ga\cr\ol\be&\ol\de\cr}}\cr
\U(1,1)=\x:
&{\scriptsize\pmatrix{\al&\be\cr\ga&\de\cr}}^\x
&={\scriptsize\pmatrix{0&1\cr 1&0\cr}}
 {\scriptsize\pmatrix{\al&\be\cr\ga&\de\cr}}^\star
{\scriptsize\pmatrix{0&1\cr 1&0\cr}}
= {\scriptsize\pmatrix{\ol\de&\ol\be\cr\ol\ga&\ol\al\cr}}\cr
\end{eq}The product of two conjugations gives a vector space automorphism,
here $\x\o\star={\scriptsize\pmatrix{0&1\cr 1&0\cr}}$.
Obviously, $\U(2)$ and $\U(1,1)$-hermiticity or unitarity are different.

\section{Three Characteristic  Time Representations}

Complex representations of the time group $\D(1)$ and the
time translations $\R$
are in unitary groups and their Lie algebras. The following three
nondecomposable types\cite{BOE,S89} are characteristic.

Irreducible  representations of time with reflection are definite unitary in $\U(1)$
\begin{eq}{rl}
\D(1)\ni e^t&\mape e^{iEt}\in\U(1)=\exp i\R\cr
\R\ni t&\mape iEt\in\log\U(1)=i\R\cr
\hbox{with}
&\hbox{eigenvalue }iE\in i\R\cr
\end{eq}These time representations are not faithful.
They are used for  stable states (particles).
As well known and repeated in the next section, the
definite unitary group leads to a  Hilbert space with
the quantum characteristic probability
amplitudes.

The smallest
faithful
nondecomposable time representations are indefinite unitary,
they are
in $\U(1,1)$
\begin{eq}{rl}
\D(1)\ni e^t&\mape
e^{iEt}{\scriptsize\pmatrix{
1&i\nu t\cr 0&1\cr}}
\in\U(1,1),~~0\ne \nu\in\R\cr
\R\ni t&\mape
i
{\scriptsize\pmatrix{
E&\nu \cr 0&E\cr}}t~~~\in\log\U(1,1)\cr
\hbox{with}&\hbox{$2\x2$-Hamilton-matrix: } H={\scriptsize\pmatrix{
E&\nu \cr 0&E\cr}}\cr
&\hbox{eigenvalue }iE\in i\R\cr
\end{eq}These  triangular reducible, but nondecomposable
 time representations with  Jordan matrices are
used for  ghosts, i.e. interactions without asymptotic states,
e.g. for the non-photonic degrees of freedom in gauge fields,
i.e. the Coulomb force and the gauge degree of freedom.
It will be discussed below in more detail.

Irreducible faithful  time representations have no time reflection
- they are valid either only for the future or for the past.
The representations of the future cone
\begin{eq}{rll}
t\ge 0:~~e^t&\mape e^{i(E+i{\Ga\over2}) t}&\in\GL(\C)=\U(1)\x\D(1),~\Ga>0\cr
t&\mape i(E+i{\Ga\over2})t&\in\log\GL(\C)=\C\cr
&\hbox{with eigenvalue }&iE- {\Ga\over2}\in\C\cr
\end{eq}with  the eigenvalues having a nontrivial real part
(width) are used for  decaying states, e.g.
for unstable particles.
The corresponding decomposable indefinite unitary
representations  are
\begin{eq}{rlccl}
\D(1)\ni e^t&\mape
e^{iEt}{\scriptsize\pmatrix{
e^{-{\Ga\over2} t}&0\cr
0&e^{+{\Ga\over2} t}\cr}}&\sim&
e^{iEt}{\scriptsize\pmatrix{
\cosh{\Ga\over2} t&\sinh{\Ga\over2} t\cr
\sinh{\Ga\over2} t&\cosh{\Ga\over2} t\cr}}&\in\U(1,1)\cr\cr
\R\ni t&\mape
{\scriptsize\pmatrix{
i(E+i{\Ga\over2})&0\cr
0&i(E-i{\Ga\over2})\cr}}t
&\sim&{\scriptsize\pmatrix{
iE&{\Ga\over2}\cr
{\Ga\over2}&iE\cr}}t&\in\log\U(1,1)
\end{eq}

\section{Stable States and Particles}

Quantum probability as used with the Hilbert space formulation of quantum
mechanics is induced  by irreducible time
representations in $\U(1)$.  The well known construction from $\U(1)$-conjugation to
 Fock-Hilbert space is shortly reviewed.

The  harmonic Bose and Fermi oscillator
give the quantum representations of the irreducible
time representation, starting from   representations
\begin{eq}{l}
t\mape \pm iEt,~~e^t\mape e^{\pm iEt}\hbox{ with }E\in\R
\end{eq}on 1-dimensional dual vector spaces with dual
bases $(\ro u,\ro u^\star)$, later called creation and annihilation operator,
related to each other by the $\U(1)$-conjugation $\star$
\begin{eq}{rl}
\U(1):&V=\C\ro u\stackrel\star\lrmap V^T=\C\ro u^\star\cr
&V\x V\map\C,~\sprod{\ro u}{\ro u}=\dprod {\ro u^\star}{ \ro u}
=1\cr
\end{eq}The time translation generator is the
basic space identity $iE\id_V$ which can be written in the
dual basic vectors (creation-annhilation operators) as tensor product
\begin{eq}{l}
iH=iE\id_V=
i E\ro u\ox\ro u^\star=
i E\rstate{\ro u}\lstate{\ro u}
\end{eq}

The noncommutative  quantum algebra\cite{S922} of the direct sum space  $V\pl V^T\cong\C^2$ -
a duality induced quotient structure of the multilinear tensor algebra
of $V\pl V^T$ -
contains as elements all  complex polynomials  $\C[\ro u,\ro u^\star]$,
modulo the  duality induced (anti)commutators
with the notation $[a,b]_\ep=ab+\ep ba$.
It is finite dimensional for Fermi (Pauli principle)
and countably infinite dimensional for Bose
\begin{eq}{rl}
\left.\begin{array}{rl}
\ep=+1&\hbox{(Fermi)}\cr
\ep=-1&\hbox{(Bose)}\cr\end{array}\right\}:&\bl Q_\ep(\C^2)=\C[\ro u,\ro u^\star]/\hbox{
modulo }\left\{\begin{array}{l}
[\ro u^\star,\ro u]_\ep-1\cr
[\ro u,\ro u]_\ep,~~[\ro u^\star,\ro u^\star]_\ep\cr\end{array}\right.\cr
&\dim_\C\bl Q_\ep(\C^2)=\left\{\begin{array}{ll}
4,&\hbox{(Fermi)}\cr
\aleph_0,&\hbox{(Bose)}\cr\end{array}\right.
\end{eq}

Therein the Hamiltonian for the $\U(1)$-time
representation above is implemented with the quantization
opposite commutator - its adjoint action gives the equations of motion
\begin{eq}{l}
\bl H=E{[\ro u,\ro u^\star]_{-\ep}\over 2}\in\bl Q_\ep(\C^2)\then
\left\{\begin{array}{ll}
[i\bl H,\ro u]&=iE\ro u\cr
[i\bl H,\ro u^\star]&=-iE\ro u^\star\cr
\end{array}\right.
\end{eq}

The Hilbert space is constructed from the quantum algebra
by canonical extension of  the scalar product $\dprod {\ro u^\star}{\ro u}=1$
for the conjugation group $\U(1)$
on the vector space $V$ to the  quantum algebra as the inner product
\begin{eq}{rll}
\bl Q_\ep(\C^2)\x \bl Q_\ep(\C^2)&\map\C,&
\sprod ab=\angle{a^\star b}\hbox{ with}\left\{\begin{array}{rl}
\angle{(\ro u^\star\ro u)^k}&=\angle{\ro u^\star\ro u}^k=1\cr
\hbox{for } k&=0,1,\dots\cr
\angle{(\ro u^\star)^k(\ro u)^l}&=0\hbox{ for }k\ne l\cr
\end{array}\right.
\end{eq}The inner product is positive semidefinite $\sprod aa\ge0$. It has
the annihilation left ideal in the quantum algebra
\begin{eq}{l}
\{n\in \bl Q_\ep(\C^2)\mid \sprod an=0\hbox{ for all }a\in\bl Q_\ep(\C^2)\}
=\bl Q_\ep(\C^2)\ro u^\star
\end{eq}The classes of the quantum algebra with respect to the annihilation
left ideal
\begin{eq}{l}
\bl Q_\ep(\C^2)\map \ro{FOCK}_\ep(\C^2),~~a\mape \rstate a = a+\bl
Q_\ep(\C^2)\ro u^\star\cr
\end{eq}constitute a complex vector space with definite scalar
product, the Fock space, 2-dimensional for Fermi
and $\aleph_0$-dimensional for Bose
\begin{eq}{l}
\ro{FOCK}_\ep(\C^2)=\bl Q_\ep(\C^2)/\bl Q_\ep(\C^2)\ro u^\star
\cong\left\{\begin{array}{ll}
\C^2,&\hbox{(Fermi)}\cr
\C^{\aleph_0},&\hbox{(Bose)}\cr\end{array}\right.\cr

\ro{FOCK}_\ep(\C^2)\x\ro{FOCK}_\ep(\C^2),~~\langle a||b\rangle=
\angle{a^\star b},~~\langle a||a\rangle=0\iff \rstate a=0\cr
\end{eq}The Fock space can be spanned by the energy eigenvectors
of the $\U(1)$-time representation as implemented in the quantum algebra
\begin{eq}{rl}
\hbox{basis of }\ro{FOCK}_\ep(\C^2):&\{\rstate k={(\ro u)^k\over\sqrt{ k!}}+\bl Q_\ep(\C^2)\ro u^\star\mid
\begin{array}{ll}
k=0,1\}&\hbox{(Fermi)}\cr
k=0,1,2\dots\}&\hbox{(Bose)}\cr\end{array}\cr
&\ro{FOCK}_\ep(\C^2)\x \ro{FOCK}_\ep(\C^2)\mape\C,~~\langle k||l\rangle =\de_{kl}
\end{eq}Its Cauchy completion defines a  Hilbert space.

In quantum mechanics, the position representation
 for the Bose case
\begin{eq}{l}
\hbox{for }\bl Q_-(\C^2):~~x={\ro u^\star+\ro u\over\sqrt2},~~
ip={d\over dx}={\ro u^\star-\ro u\over\sqrt2},~~
\bl H=E{\acom{\ro u}{\ro u^\star}\over 2}=E{p^2+x^2\over2}

\end{eq}gives the square
integrable functions $\ol{\ro{FOCK}}_-(\C^2)\cong L^2_{dx}(\R,\C)$
with the Hermite polynomials $\ro H_k$
\begin{eq}{l}
{\ro u^k\over \sqrt{k!}}\rstate0=\rstate k\cong\psi_k:~~
\left\{\begin{array}{rl}
\ro u^\star\rstate 0&=0\then (x+{d\over dx})\psi_0(x)=0\cr
\psi_k(x)&={1\over\sqrt{k!}}
\({x-{d\over dx}\over\sqrt2}\)^k\psi_0(x)
={1\over\sqrt{ 2^kk!\sqrt\pi}} e^{{x^2\over2}}
\(-{d\over dx}\)^ke^{-x^2}\cr
&={1\over\sqrt{ 2^kk!\sqrt\pi}} e^{-{x^2\over2}}\ro H_k(x)\cr\end{array}\right.
\end{eq}

All particle quantum fields are built with harmonic oscillators
where the creation and annihilation operators are indexed with
momenta $\rvec q\in\R^3$
\begin{eq}{l}
 V_{\rvec q}=\C \ro u(\rvec q)\stackrel\star\lrmap
 V^T_{\rvec q}=\C \ro u^\star(\rvec q)~~\hbox{ with }
\left\{\begin{array}{ll}
[\ro u^\star(\rvec p),\ro u(\rvec q)]_\ep&=(2\pi)^3\de(\rvec q-\rvec p)\cr
[\ro u(\rvec p),\ro u(\rvec q)]_\ep&=0\cr
[\ro u^\star(\rvec p),\ro u^\star(\rvec q)]_\ep&=0\cr
\angle {\ro u^\star(\rvec p)\ro u(\rvec q)}&=(2\pi)^3\de(\rvec q-\rvec p)\cr

\end{array}\right.
\end{eq}e.g.    for a  stable spinless    $\pi^0$
with a Lorentz scalar field $\bl \Phi$,
 for a  stable spin 1 $Z^0$ with a Lorentz vector  field $\bl Z$
 or for the spin ${1\over2}$ electron-positron
with particles $(\ro u,\ro u^\star)$
 and antiparticles  $(\ro a,\ro a^\star)$ in the left and right handed contributions
with a Dirac field $\bl\Psi=(\bl r^A,\bl l^{\dot A})$ - all given by
direct integrals $\int^{\hskip-1mm \pl}$ over the
$\rvec q$-indexed subspaces
 \begin{eq}{rllll}
\bl\Phi(x)&
=\int^{\hskip-1mm \pl}{d^3q\over (2\pi)^3}&&
{  \ro u(\rvec q)e^{iqx}+
 \ro u^\star(\rvec q)e^{-iqx}\over\sqrt2}\cr
\bl Z^j(x)
&=\int^{\hskip-1mm \pl}{d^3q\over (2\pi)^3}&
\La({q\over m})^j_a&
{  \ro u^a(\rvec q)e^{iqx}+ \ro u^{\star a}(\rvec q)e^{-iqx}\over\sqrt2},&
\left\{\begin{array}{rl}
j&=0,1,2,3\cr
a&=1,2,3\cr\end{array}\right.
\cr
\bl r^{A}(x)
&= \int^{\hskip-1mm \pl}{d^3q\over (2\pi)^3}
& s({q\over m})^A_\al&{ \ro u^\al(\rvec q)e^{iqx}+
\ro a^{\star\al}(\rvec q) e^{-iqx}\over\sqrt2}
 \cr
\bl l^{\dot A}(x)
&=\int^{\hskip-1mm \pl}{d^3q\over (2\pi)^3}
&s^{-1}({q\over m})_\al^{\dot A}
&{ \ro u^\al(\rvec q)e^{iqx}- \ro a^{\star \al}(\rvec q)e^{-iqx}\over\sqrt2}
,&\left\{\begin{array}{rl}
A,\dot A&=1,2\cr
\al&=1,2\cr\end{array}
\right.
\cr
\end{eq}The nonscalar fields involve the corresponding
re\-pre\-sen\-ta\-tions of
the boosts $\SO_0(1,3)/\SO(3)\cong\SL(\C^2)/\SU(2)$  - here the
Weyl and vector representations
$\{s({q\over m}),\La({q\over m})\}$.

\section{Ghosts without Particles}
 Nondiagonalizable
time representation in the indefinite unitary group $\U(1,1)$ are used for
the nonparticle ghost degrees of freedom in gauge and Fadeev-Popov fields.
The indefinite unitary time representation will be considered, first in
an algebraic matrix model, then in the associated quantum algebras
and, finally,  in the relativistic gauge fields.

\subsection{Indefinite Unitarity for Ghosts}

The dynamics of a   Newtonian free mass point
with Hamiltonian $H={p^2\over 2M}$
\begin{eq}{rl}
{\scriptsize \pmatrix{ x\cr
i p\cr}}(t)&=
{ \scriptsize\pmatrix{
1&-i{t\over M}\cr 0&1\cr}}
{\scriptsize \pmatrix{ x\cr
i p\cr}}\iff\left\{\begin{array}{rl}
x(t)&=x(0)+{t\over M}p(0)\cr
p(t)&=p(0)\cr\end{array}\right.\cr
{d\over dt}
{\scriptsize \pmatrix{ x\cr
i p\cr}}&=
i{ \scriptsize\pmatrix{
0&-{1\over M}\cr 0&0\cr}}
{\scriptsize \pmatrix{ x\cr
i p\cr}}
\end{eq}is a faithful nondiagonalizable
time representation, in general
\begin{eq}{rl}
e^t&\mape
e^{iEt}{\scriptsize \pmatrix{
1&i\nu t\cr 0&1\cr}}\hbox{ with }E,\nu\in\R,~~\nu\ne 0\cr
t&\mape
i
{\scriptsize \pmatrix{
E&\nu \cr 0&E\cr}}t
\end{eq}For the Newtonian mass point momentum is an eigenvector
with trivial time translation eigenvalue, position is  a nilvector
(principal vector, no eigenvector).

The Hamilton-matrix is $\U(1,1)$-hermitian
and the group representation $\U(1,1)$-unitary (1st section)
\begin{eq}{rlll}
 H&=&{\scriptsize\pmatrix{
E&\nu \cr 0&E\cr}}&= H^\x\cr
D(t)&=e^{iEt}&{\scriptsize\pmatrix{
1&i\nu t\cr 0&1\cr}}&= D(-t)^\x\cr
\end{eq}The Hamilton-matrix is the sum of two commuting transformations,
the  semisimple
and the  nilpotent part, called  nil-Hamiltonian
\begin{eq}{l}
H=E\bl1_2+\nu N, ~~ N={\scriptsize\pmatrix{
0&1 \cr 0&0\cr}},~~
[ H, N]=0,~~ N^2=0
\end{eq}

The representation space
cannot be spanned by energy eigenvectors alone which are
characterized by the trivial action of the nil-Hamiltonian
\begin{eq}{l}
\begin{array}{c}
\hbox{eigenvector (`good')}\cr
\rstate{\ro G}=
{\scriptsize\pmatrix{0\cr 1\cr}},~~
\lstate{\ro  G}=(1,0)\end{array},\left\{\begin{array}{rl}
H\rstate{\ro G}&=E\rstate{\ro G},~~ N\rstate{\ro G}=0\cr
(H-E)\rstate{\ro G}&=0\cr
\rstate{\ro G}(t)&=e^{iEt}\rstate{\ro G}\cr\end{array}\right.
\end{eq}In addition to the eigenvector one needs another
principal vector
\begin{eq}{l}
\begin{array}{c}
\hbox{nilvector (`bad')}\cr
\rstate{\ro B}={\scriptsize\pmatrix{1\cr 0\cr}},~~
\lstate{\ro B}=(0,1)\end{array},
\left\{\begin{array}{rl}
 H\rstate{\ro B}&=E\rstate{\ro B}+\nu\rstate{\ro G},
 ~N\rstate{\ro B}=\nu\rstate{\ro G}\cr
(H-E)^2\rstate{\ro B}&=0\cr
\rstate{\ro B}(t)&=e^{iEt}\rstate{\ro B}+i\nu te^{iEt}\rstate{\ro G}\cr
\end{array}\right.
\end{eq}For the $\U(1,1)$-inner product both eigenvectors and nilvectors are
 ghosts, i.e. their $\U(1,1)$-norm vanishes in Witt bases
\begin{eq}{rl}
\U(1,1):&
{\scriptsize\pmatrix{0&1\cr 1&0\cr}}
\then\left\{\begin{array}{ll}
\sprod{\ro G}{\ro G}=0,&\sprod{\ro G}{\ro B}=1\cr
\sprod{\ro B}{\ro G}=1,&
\sprod{\ro B}{\ro B}=0\cr\end{array}\right.\cr
&H=E\(\rstate{\ro B}\lstate{\ro G}+\rstate{\ro G}\lstate{\ro B}\)+\nu
\rstate{ \ro G}\lstate{\ro G}\cr
\end{eq}Eigen- and nilvectors are not $\U(1,1)$-orthogonal.
Obviously the inner product gives no Hilbert space structure
as seen in Sylvester bases
\begin{eq}{l}
{\scriptsize\pmatrix{0&1\cr 1&0\cr}}
\sim {\scriptsize\pmatrix{1&0\cr 0&-1\cr}}
\then\left\{\begin{array}{ll}
\sprod{\ro G+\ro B}{\ro G+\ro B}=2,&
\sprod{\ro G+\ro B}{\ro G-\ro B}=0\cr
\sprod{\ro G-\ro B}{\ro G+\ro B}=0,&
\sprod{\ro G-\ro B}{\ro G-\ro B}=-2\cr
\end{array}\right.
\end{eq}

\subsection{Ghosts in Quantum Structures}

The quantum structure of the $\U(1,1)$-time representation in the
last subsection becomes somewhat more complicated
since the nilpotency $N^2=0$ for the matrix product
(endomorphism product) cannot
be implemented with the quantum algebra product
using  Bose degrees of freedom alone.
This requires the
introduction of twin-like Fermi
degrees of freedom as done with the Fadeev-Popov\cite{KO}  fields
as Fermi twins for the nonparticle degrees of freedom in the Bose gauge
fields.

By using a doubling of the basic
representation space and constructing a $\Z_2$-graded
quantum algebra (last section) with both a  Bose  (capital letters $\ro G,\ro B$)
 and a Fermi (small letters $\ro g,\ro b$) factor
 \begin{eq}{rl}
\bl Q_-(\C^4)\ox \bl Q_+(\C^4)&\cong
\C[\ro B,\ro G,\ro B^\x,\ro G^\x]
\ox\C[\ro b,\ro g,\ro b^\x,\ro g^\x]\cr
&\hbox{with }
\left\{\begin{array}{rr}
\com{\ro G^\x}{\ro B}=1,&
\com{\ro B^\x}{\ro G}=1\cr
\acom{\ro g^\x}{\ro b}=1,&
\acom{\ro b^\x}{\ro g}=1\cr
\end{array}\right.\cr
\end{eq}the time development is implemented by the doubled Hamiltonian
\begin{eq}{l}

\bl H_{BF}=\bl H_B+\bl H_F,~~\left\{\begin{array}{rl}
\bl H_B&= E {
\acom{\ro B}{\ro G^\x}+\acom{\ro G}{\ro B^\x}\over 2}+\nu \ro G\ro G^\x\cr
\bl H_F&= E {
\com{\ro b}{\ro g^\x}+\com{\ro g}{\ro b^\x}\over 2}+\nu \ro g\ro g^\x\cr
\end{array}\right.\cr
\end{eq}By products of Bose with Fermi operators
a nilquadratic BRS-operator\cite{BRS,S911} of Fermi type  can be constructed
\begin{eq}{l}
\bl N_{BF}=\ro g\ro G^\x+\ro G\ro g^\x\then [\bl H_{BF},\bl N_{BF}]=0,~~ \bl N_{BF}^2=0\cr
\end{eq}The quantum product $\bl N_B=\ro G\ro G^\x$ of Bose operators
is not nilpotent.

In the basic vector space formulation there arises
as Hamilton-matrix and  BRS matrix
\begin{eq}{rl}
 H_{BF}= H_B\pl H_F&={\scriptsize\pmatrix{ H&0\cr0& H\cr}}={\scriptsize
\left(\begin{array}{cc|cc}
E &\nu&0&0\cr
0& E &0&0\cr\hline
0&0& E &\nu\cr
0&0&0&E \cr
\end{array}\right)}\cr

 N_{BF}&={\scriptsize\pmatrix{0& N\cr N&0\cr}}={\scriptsize
\left(\begin{array}{cc|cc}
0&0&0&1\cr
0&0&0&0\cr\hline
0&1&0&0\cr
0&0&0&0\cr\end{array}\right)}
\end{eq}

The BRS-operator effects  - in analogy to the nil-Hamiltonian action in the
$2\x2$-matrix formulation -
the projection
to a  subspace, spanned by time
translation eigenvectors.
The graded  adjoint action with the  BRS-operator -
replacing the classical gauge transformation
\begin{eq}{l}
\ad\bl N_{BF} (a)=\left\{\begin{array}{lll}
[\bl N_{BF},a]&\hbox{for $a$ Bose},&\hbox{e.g. }[\bl N_{BF},\ro G]=0\cr
\acom{\bl N_{BF}}a&\hbox{for $a$ Fermi},&\hbox{e.g. }\acom{\bl N_{BF}}{\ro g}=0
\cr\end{array}\right.\cr

\end{eq}defines the unital subalgebra with
the  linear combinations of the time translation eigenvectors
- replacing the gauge invariant operators
\begin{eq}{l}
\INV_{\bl N_{BF}}\bl Q_\pm(\C^8)=\{p\in
\bl Q_-(\C^4)\ox \bl Q_+(\C^4)
\mid \ad\bl N_{BF}(p)=0\}
\end{eq}

The product Fock space for $\bl Q_-(\C^4)\ox \bl Q_+(\C^4)$,
as constructed for the $\U(1)$-time representations in the last section,
has an indefinite metric
\begin{eq}{l}
\ro{FOCK}_-(\C^2)\ox \ro{FOCK}_+(\C^2)
\hbox{ with }\left\{\begin{array}{ll}
\sprod{\ro G\pm \ro B}{\ro G\pm \ro B}&=\pm 2\cr
 \sprod{\ro g\pm \ro b}{\ro g\pm \ro b}&=\pm 2\end{array}\right.
\end{eq}The subspace
with the time translation eigenvectors - i.e. the
classes for the BRS-invariance algebra $\INV_{\bl N_{BF}}\bl Q_\pm(\C^8)$
above, replacing the gauge invariant states
\begin{eq}{l}
\{\rstate p\in \ro{FOCK}_-(\C^2)\ox \ro{FOCK}_+(\C^2)
\mid \bl N_{BF}\rstate p=0\}
\end{eq}contains - up to $\rstate 0$ (the class of the quantum
algebra unit $1$)
with $\sprod 00=1$ -
only normless vectors (ghosts),
e.g. $\sprod {\ro g}{\ro g}=0=\sprod {\ro G}{\ro G}$.
Its metric is semidefinite. The associated
Hilbert space with the definite classes is 1-dimensional $\C\rstate 0$,
i.e. it contains  only the classes of the scalars.
From the whole operator quantum algebra for
the $\U(1,1)$-time representation there is -
apart form the scalars $\C\rstate 0$ -
no vector left for the asymptotic Hilbert space - ghosts have no asymptotic
states, i.e. they have no particle projections.

\subsection{Ghosts in Gauge Theories}

$\U(1,1)$-time representations with the characteristic ghost pairs
$(\ro B,\ro G)$ arise in
gauge fields which embed via the spacetime translation representations
the indefinite Lorentz group into  an indefinite unitary
groupt $\SO_0(1,3)\subnoteq \U(1,3)$.
Orthogonal time-space bases (Sylvester bases)
and lightlike bases (Witt bases)
reflect the two bases for eigen- and nilvector used above
\begin{eq}{l}
{\scriptsize\pmatrix{1&0\cr 0&-1\cr}}
\sim{\scriptsize\pmatrix{0&1\cr 1&0\cr}},~~
x_0^2-x_3^2=(x_0+x_3)(x_0-x_3)
\end{eq}

The algebraic concepts used in the $(2\x 2)$-matrix language  above
are the left hand side of  the following dictionary
 for the gauge field language
\begin{eq}{rcl}
\hbox{nilconstant $\nu\ne0$}&\sim&\hbox{gauge fixing constant}\cr
\hbox{nil-Hamiltonian $N$ with $ N^2=0$}&\sim&\hbox{Becchi-Rouet-Stora charge}\cr
\hbox{nil-Hamiltonian action, e.g. $N\rstate{\ro B}=\nu\rstate{\ro G}$}
&\sim&\hbox{gauge transformation}\cr
\hbox{time translation eigenvectors, e.g. $ N\rstate{\ro G}=0$} &\sim&\hbox{gauge invariant vectors}\cr
\hbox{eigenvectors with nontrivial norm} &\sim&\hbox{asymptotic particles}\cr
\hbox{ghost pairs with trivial norm} &\sim&\hbox{interaction without
particles}\cr
\end{eq}

For a free massless electromagnetic gauge field the quantization
\begin{eq}{l}
\com {\bl A^k}{\bl A^j}(x)
=\int{d^4q\over(2\pi)^3}\ep(q_0)[
 -\eta^{kj}+2\nu q^kq^j{\p\over\p q^2}]\de(q^2)
 e^{iqx}  \cr
\end{eq}is contrasted
with the quantization of a  free massive vector field
\begin{eq}{l}
\com {\bl Z^k}{\bl Z^j}(x)
=\int{d^4q\over(2\pi)^3}\ep(q_0)[-\eta^{kj}+{ q^kq^j\over q^2}]
\de(q^2-m^2) e^{iqx},~~m^2>0  \cr
\end{eq}The gauge field employs the characteristic
Dirac function derivative $\de'(q^2)$, multiplied with the gauge fixing constant
$\nu$.

The harmonic analysis of the massive vector field with respect to
the time representations with spin 1 involves the Lorentz transformation
$\La({q\over m})$ to a rest system with $\SO(3)$ the `little group' for
energy-momenta with $q^2=m^2>0$
\begin{eq}{rl}
\bl Z^j(x)
&=\int^{\hskip-2mm}{d^3q\over (2\pi)^3}
\La({q\over m})^j_a
{  \ro u^a(\rvec q)e^{iqx}+ \ro u^{\star a}(\rvec q)e^{-iqx}\over\sqrt2}
\hbox{ with }q_0=\sqrt{m^2+\rvec q^2}
\cr
\hbox{and }\La({q\over m})
&= {1\over m}{\scriptsize \left(\begin{array}{c|c}
 q_0 &\rvec q\cr\hline
\rvec q&\de_{ab}m+{q_aq_b\over  m+q_0}\cr\end{array}\right)}\in\SO_0(1,3),~~
\La({q\over m}){\scriptsize\pmatrix{m\cr0\cr0\cr0\cr}}=q,~~
\end{eq}The time representations in the gauge field
\begin{eq}{rl}
\bl A^j(x) &=\int{d^3 q\over (2\pi)^3}~h({\rvec q\over |\rvec q|})^j{1\over\sqrt{q_0}}
{\pmatrix{
{[\ro B(\rvec q)-i\nu  x_0q_0\ro G(\rvec q)] e^{iqx}
+(1+\nu)\ro G^\x(\rvec q) e^{-iqx}\over\sqrt2}\cr
{\ro u^1(\rvec q) e^{iqx}+\ro u_1^{\star}(\rvec q) e^{-iqx}\over\sqrt2}\cr
{\ro u^2(\rvec q) e^{iqx}+\ro u_2^\star(\rvec q) e^{-iqx}\over\sqrt2}\cr
{(1+\nu)\ro G(\rvec q) e^{iqx}
+[\ro B^\x(\rvec q)+i\nu  x_0q_0\ro G^\x(\rvec q)] e^{-iqx}\over\sqrt2}\cr}}
\cr
&\hbox{with }q_0=|\rvec q|\cr
\end{eq}involve - in addition to the two
photonic particle degrees of freedom in the 1st and 2nd component
with two time representations in $\U(1)$
- the Coulomb interaction and gauge degree
of freedom in the 0th and 3rd component.  $h({\rvec q\over |\rvec q|})$
is a representative of $\SO_0(1,3)/\SO(2)$
to transform to the polarization group $\SO(2)$ (`little group' for
energy-momenta with $q^2=0$, $q\ne0$)
\begin{eq}{l}
h({\rvec q\over |\rvec q|})
=
{\scriptsize \left(
\begin{array}{c|c}
1&0\cr\hline
0&O({\rvec q\over |\rvec q|})\cr
\end{array}\right)}\o w,~~

w={1\over\sqrt2}{\scriptsize\left(\begin{array}{c|c|c}
1&0&-1\cr\hline
0&\bl1_2\sqrt2&0\cr \hline
1&0&1\cr\end{array}\right)}\cr
O({\rvec q\over |\rvec q|})\in\SO(3)
\hbox{ with }O({\rvec q\over |\rvec q|}){\scriptsize\pmatrix{0\cr0\cr|\rvec q|\cr}}=\rvec q

\end{eq}

The $\U(1,1)$-time representation structure of the Faddev-Popov
Fermi scalar fields is similar
and will not be given here explicitly.

The inner product structure  can
be seen in the decomposition of the indefinite time representation containing
unitary group $\U(1,3)$
extending the Lorentz group - for massive vector fields with  Sylvester
bases
\begin{eq}{l}
\left.\begin{array}{c}
\hbox{massive particles,}\cr
\hbox{e.g. stable $Z$:}\end{array}\right\}
\hbox{ with }\left\{\begin{array}{l}
\SO_0(1,3)\inmap\U(1,3)\supnoteq \U(1)\x\U(3)\cr
\hbox{Lorentz metric: }
{\scriptsize\left(\begin{array}{c|c}
-1&0\cr\hline 0&\bl1_3\cr\end{array}\right)}=-\eta\end{array}\right.\cr
\end{eq}and for massless gauge fields
with Witt bases
\begin{eq}{l}
\left.\begin{array}{c}
\hbox{massless ghosts and}\cr
\hbox{particles, e.g. photons $\ga$:}\end{array}\right\}\hbox{ with }
\left\{\begin{array}{l}
\SO_0(1,3)\inmap\U(1,3)\supnoteq \U(1,1)\x\U(2)\cr
\hbox{Lorentz metric: }\cr
\hfill{\scriptsize\left(\begin{array}{c|c|c}
0&0&1\cr\hline
0&\bl1_2&0\cr\hline
1&0&0\cr\end{array}\right)}=-
w^T\o \eta \o w
\end{array}\right.\cr
\end{eq}

\section{Unstable States and Particles}

Decaying states (particles) can be considered in a Hilbert space where
they form, together with other states - stable or unstable - multidimensional
probability collectives.
The 2-dimensional neutral
kaon system with the short and long lived
unstable neutral kaon as an illustration
leads to the general algebraic formulation.

\subsection{The Neutral Kaons as a Probability Collective}

The system of the two neutral $K$-meson states
$\rstate {K  _{S,L}}$
shows  - on the one hand - the phenomenon of CP-violation
(treated in this subsection) and - on the other hand - is unstable
and decays into many channels (treated with the general formlism
in the next subsection).

The kaon particles
are no  CP-eigenstates  $\rstate{K_{\pm}}$ to which they can be transformed by
an invertible $(2\x 2)$-matrix
\begin{eq}{rl}
{\scriptsize\pmatrix{\rstate {K_S  }\cr\rstate {K_L  }\cr}}
&=T{\scriptsize\pmatrix{
\rstate {K_+}\cr\rstate {K_-}\cr}},~~T\in\GL(\C^2)
\end{eq}The CP-eigenstates are fictive in the sense that there are no
observable particles connected with them.
Under the assumption of CPT-invariance
the matrix is symmetric and parame\-tri\-zable by
two complex numbers
wherefrom - with irrelevant $\U(1)$-phases - the normalization
$N_K$ can be chosen to be real
\begin{eq}{l}
T=T^T={1\over N_K\sqrt{1+|\ep|^2}}
{\scriptsize\pmatrix{1&\ep\cr \ep&1\cr}}
,~~\ep\in\C,~N_K\in \R\cr
\end{eq}

The time development is implemented by a Hamiltonian, non-hermitian for
the unstable states $H_K\ne H_K^\star$
\begin{eq}{l}
\hbox{for }t\ge0:~
{d\over dt}
{\scriptsize\pmatrix{\rstate {K_+}\cr\rstate {K_-}\cr}}=iH_K
{\scriptsize\pmatrix{\rstate {K_+}\cr\rstate {K_-}\cr}},~~
{d\over dt}{\scriptsize\pmatrix{
\rstate {K_S  }\cr\rstate {K_L  }\cr}}=i\diag H_K
{\scriptsize\pmatrix{\rstate {K_S  }\cr\rstate {K_L  }\cr}}\cr
\end{eq}with the diagonal form for the energy eigenstates
\begin{eq}{rl}
\diag H_K&=
{\scriptsize\pmatrix{M_S&0\cr 0&M_L\cr}},~~
M=m+i{\Ga\over2},~m,~\Ga>0\cr
H_K&=T^{-1}\diag H_K~T={\scriptsize\pmatrix{M_S-\ep^2M_L&\ep(M_S-M_L)\cr
\ep(M_L-M_S)&M_L-\ep^2 M_S\cr}}\cr
\end{eq}

The scalar product, constructed for $t=0$,  is time independent.
The CP-eigenstates  constitute an orthogonal basis in the
complex 2-dimensional Hilbert space
\begin{eq}{l}
\hbox{CP-eigenstates: }
{\scriptsize\pmatrix{
\sprod{K_+}{K_+}&\sprod{K_+}{K_-}\cr
\sprod{K_-}{K_+}&\sprod{K_-}{K_-}\cr}}
={\scriptsize\pmatrix{1&0\cr0&1\cr}}\cr
\end{eq}wherefrom there arises the positive\footnote{\scriptsize Any matrix product $f^\star f$
is unitarily equivalent to a positive diagonal matrix.}
 non-diagonal
 $T^\star T$ for the  time translation  eigenstates
\begin{eq}{l}
\hbox{energy eigenstates: }
{\scriptsize\pmatrix{
\sprod{K_S  }{K_S  }&\sprod{K_S  }{K_L  }\cr
\sprod{K_L  }{K_S  }&\sprod{K_L  }{K_L  }\cr}}
=T^\star T
=(T^\star T)^\star
={1\over N_K^2}
{\scriptsize\pmatrix{1&\de\cr
\de &1\cr}}\cr
\end{eq}The experiments give a nontrivial
transition between the short and long lived kaon - the real part of $\ep$.
Therefore $T$ is not definite unitary
\begin{eq}{l}
 \de={\ep+\ol\ep\over 1+|\ep|^2}\sim 0.327\x10^{-2}\then T\notin\U(2)
\end{eq}

The normalization $N_K$, usually chosen, normalizes individually
the particles states, i.e. vectors with
$\sprod{K_S  }{K_S  }=\sprod {K_L  }{K_L  }=1$.
The kaon  system is a
2-dimensional probability collective, i.e. a complex
2-dimensional Hilbert space.
Therefore it will be
collectively normalized via the discriminant (determinant)
\begin{eq}{rl}
\det T^\star T=|\det T|^2=\sprod {\det T}{\det T}&=\sprod{K_S  }{K_S  }\sprod{{K_L  }}{K_L  }-
|\sprod{ {K_S  }}{K_L  }|^2
=1\cr
\then N_K^2&={\sqrt{ (1-\ep^2)(1-\ol\ep^2)}\over 1+|\ep|^2}= 1-\de^2
\end{eq}

Since the energy eigenstates are not orthogonal, i.e. the
transformation $T$ is not unitary
 $T^\star\ne T^{-1}$, there exists a 3rd basis
 of the 2-dimensional Hilbert space
 \begin{eq}{l}
{\scriptsize\pmatrix{\rstate {K_S^\perp}\cr\rstate {K_L^\perp}\cr}}=
T^{-1\star}{\scriptsize\pmatrix{
\rstate {K_+}\cr\rstate {K_-}\cr}}
=(TT^\star )^{-1}{\scriptsize\pmatrix{\rstate {K_S  }\cr\rstate {K_L  }\cr}}
\end{eq}also not orthogonal,
 but orthogonal  with the energy eigenstates
\begin{eq}{l}
{\scriptsize\pmatrix{
\sprod{K_S^\perp}{K_S^\perp}&\sprod{K_S^\perp}{K^\perp_L}\cr
\sprod{K_L^\perp}{K_S^\perp}&\sprod{K^\perp_L}{K^\perp_L}\cr}}=
(T^\star T)^{-1},~~
{\scriptsize\pmatrix{
\sprod{K_S^\perp}{K_S  }&\sprod{K_S^\perp}{K_L  }\cr
\sprod{K_L^\perp}{K_S  }&\sprod{K^\perp_L}{K_L  }\cr}}=
{\scriptsize\pmatrix{1&0\cr0&1\cr}}
\end{eq}

A decomposition of the unit can be written with
the orthogonal and non-orthogonal  bases
\begin{eq}{rl}
\bl 1_2
&=\rstate{K_+}\lstate {K_+}+
\rstate{K_-}\lstate {K_-}\cr&
=\rstate{K_S  }\lstate {K_S^\perp}+\rstate{K_L  }\lstate {K_L^\perp}
=\rstate {K_S^\perp}\lstate{K_S  }+\rstate {K_L^\perp}\lstate{K_L  }\cr
&=\rstate  {K_S  } \lstate{K_S  }-\de \rstate {K_S  }\lstate{K_L  }
-\de \rstate {K_L  }\lstate{K_S  }+\rstate  {K_L  }\lstate{K_L  }
\hbox{ for }{N_K^2\over  1-\de^2}=1\cr
\end{eq}

\subsection{Non-Orthogonal Decaying States}

The possibility to have nontrivial transition elements between
particles,
as $\sprod {K_S}{K_L}$ above,
can be connected to  the deviation from the definite unitary structures for
unstable states. The following well known  theorems with respect to unitary
equivalence are
relevant for the situation.

An $(n\x n)$-matrix $H$
acting on a vector space $V\cong\C^n$, e.g. a Hamilton-matrix
for the time translations, is unitarily equivalent to a  diagonal matrix
if,  and only if it is normal - all concepts with respect
to a definite $\U(n)$-conjugation
\begin{eq}{l}
H=U\o\diag H\o U^\star \hbox{ with }U\in\U(n)
\iff H\o H^\star=H^\star\o H
\end{eq}In this case, the vector space can be decomposed as
orthogonal sum of the eigenspaces for
$N$ different eigenvalues $\spec H=\{M_k\}$
\begin{eq}{l}
V={\PL_{k=1}^N} V_k,~~
\diag H={\PL_{k=1}^N}  M_k\id_{V_k},~~ \sprod{V_k}{V_l}=\{0\}
\hbox{ for }M_k\ne M_l
\end{eq}For $\U(n)$-hermitian operators $H=H^\star$ the eigenvalues are real
$M=E\in\R$.

An analogue (real)  diagonal structure holds for
(selfadjoint) normal operators on infinite dimensional Hilbert spaces.

Hamiltonians acting on a Hilbert space  with
complex eigenvalues $E+i{\Ga\over 2}$, $\Ga>0$,
have to be $\U(n)$-nonhermitian, $H\ne H^\star$. Only with at least one
 non-real energy involved, i.e. one unstable particle,
 two time translation eigenvectors (particles) with different
energies
can have a nontrivial transition element - unstable particles
have not  to be
orthogonal to other particles.

This structure raises a basic question,
dicussed already in the 1st section: A Hamiltonian implements the real
time translations as acting upon a complex vector space. To recognize the realness also
in the complex structure by $H=H^*$,
 the complex vector space has to come with a
conjugation
or - equivalently - with an inner product, characterized by a unitary invariance
group. The conjugation above for the probability structure has to
be definite with
invariance group $\U(n)$. Possibly, the real spacetime
translations for unstable states with complex eigenvalues
are implemented by
operators which are hermitian under an
indefinite unitary group where the definite probability group comes as a
subgroup. The Lorentz embedding group
$\U(1,3)$ with probability subgroups $\U(3)$
for massive and $\U(2)$ for massless particles has been mentioned above.
Another example is the Dirac spinor conjugation group $\U(2,2)$
as familiar from the left-right handed conjugation.
The embedded definite $\U(2)$-conjugation is used for the probability structure,
the indefinite one $\U(1,1)$ for the particle-antiparticle reflection.
The indefinite unitary structures for unstable states in general are not
discussed in this paper.

\subsection{Probability Collectives for Decaying Particles}

The two translation eigenstates (particles)
for  unstable kaons are generalized to $q$ eigenstates
$\rstate {M}$
(particles) with the eigenvalues $M=m+i{\Ga\over 2}$
involving at least
one unstable state $\Ga>0$. An orthogonal basis,
related to $\rstate M$ by a non-unitary transformation $T\notin \U(q)$
is denoted
by $\rstate U$ - generalizing the CP-eigenstates of the kaon collective.
In addition
 all stable decay modes, assumed to be $p$ translation eigenstates
 $\rstate E$ (particles) with real eigenvalues $E$ are included, e.g.
$\rstate{\pi,\pi}$, $\rstate{\pi,\pi,\pi}$,
$\rstate{\pi,l,\nu_l}$ for the kaon collective
\begin{eq}{rll}
\rstate {{M}}&=
\(\rstate{m_j+i{\Ga_j\over 2}}\)_{j=1}^q&
\left\{\begin{array}{l}
\hbox{eigenstates with at least}\cr
\hbox{one decaying channel}\end{array}\right.
\cr\cr
\rstate {U}&=
\(\rstate{U_j}\)_{j=1}^q&\hskip5mm\hbox{orthogonal states }
\cr\cr
\rstate E&=
\(\rstate{E_l}\)_{l=1}^p&\left\{\begin{array}{l}
\hbox{stable eigenstates}\cr
\hbox{(decay channels)}\end{array}\right.
\cr
\end{eq}In a more
 general formulation also an infinite dimensional
momentum dependence can be included.

The translation eigenstates
have the time development with a diagonal
Hamiltonian
\begin{eq}{l}
\hbox{for }t\ge0:~~
{d\over dt}
{\scriptsize\pmatrix{\rstate {U}\cr\rstate {E}\cr}}=
iH
{\scriptsize\pmatrix{\rstate {U}\cr\rstate {E}\cr}},~~
WHW^{-1}=\diag H={\scriptsize\pmatrix{{M}&0\cr 0&E\cr}}
\end{eq}The
equivalence transformation $W\notin\U(q+p)$
is the product of a triangular matrix
with  a $(p \x
q)$-matrix $w$ from the  decay $\rstate{M}\to \rstate E$,
called Wigner-Weisskopf\cite{WIWEI} matrix,
and a matrix with a
$(q \x q)$-matrix $T$  for the
transformation $\rstate M=T\rstate U$
\begin{eq}{rl}
W&={1\over N}{\scriptsize\pmatrix
{\bl1_q&w\cr 0&\bl1_p\cr}\pmatrix{T&0\cr 0&\bl1_p\cr}}=
{1\over N}{\scriptsize\pmatrix{T&w\cr 0&\bl1_p\cr}}\cr
\cr
H&={\scriptsize\pmatrix{T^{-1}&0\cr 0&\bl1_p\cr}}
{\scriptsize\pmatrix{{M}&
w({M}-E)\cr 0&E\cr}}{\scriptsize\pmatrix{T&0\cr 0&\bl1_p\cr}}
={\scriptsize\pmatrix{H_U&T^{-1}w({M}-E)\cr 0&E\cr}}
\end{eq}

The eigenstates
$\rstate {{M}}$
have projections both on the orthogonal states and on the decay channels
\begin{eq}{l}
\hbox{eigenstates: }
{\scriptsize\pmatrix{\rstate {{M}}\cr\rstate {E}\cr}}={1\over N}
{\scriptsize\pmatrix{T&w\cr 0&\bl1_p\cr}}
{\scriptsize\pmatrix{\rstate {U}\cr\rstate {E}\cr}}
={1\over N}{\scriptsize\pmatrix
{T\rstate {U}+w\rstate {E}\cr\rstate {E}\cr}}\cr
\end{eq}

The scalar product matrix for the probability collective with the
$(q+p)$ translation eigenstates
 arises from the diagonal matrix with the orthogonal states
and the decay channels
\begin{eq}{l}
{\scriptsize\pmatrix{
\sprod{U}{U}&\sprod{ U}{E}\cr
\sprod{ E}{U}&\sprod{ E}{E}\cr}}={\scriptsize\pmatrix{
\bl1_p&0\cr 0&\bl1_q\cr}},~~
{\scriptsize\pmatrix{
\sprod{{M}}{{M}}&\sprod{{M}}{E}\cr
\sprod{ E}{{M}}&\sprod{ E}{E}\cr}}
=W^\star W={1\over N^2}{\scriptsize\pmatrix{
T^\star T&T^\star w\cr w^\star T& \bl1_p+w^\star w\cr}}
\end{eq}with the collective  normalization
\begin{eq}{rl}
\sprod{\det W}{\det W}
&=\sprod{M}{M}\sprod{ E}{E}-\sprod{ M}{E}\sprod{ E}{M}\cr

&=\det T^\star T~[\det (\bl1_p+w^\star w)-\det w^\star w]=N^2=1
\end{eq}

\subsection{Probabilities with Nonabelian Groups $\U(n)$}

The scalar product $\sprod ~~$ of the complex space $V\cong\C^n$, $n=q+p$,
with the translation eigenstates (particles) involving at least one unstable
state
is a positive matrix, not the unit matrix. It can be factorized
with a nonunitary matrix $W\notin\U(n)$ chosen\footnote{\scriptsize
The literally wrong notation $g\in G/H$
denotes a coset representative, correct:
$g\in gH\in G/H$.}
from the
orientation manifold $\GL(\C^n)/\U(n)$
\begin{eq}{l}
\bl1_n\ne \sprod ~~=W^\star W,~~W\in \GL(\C^n)/\U(n)
\end{eq}The  individual probability normalization
for one state by the scalar product
\begin{eq}{l}
\hbox{for }\U(1):~~\sprod {\ro u}{\ro u}=1
\end{eq}is generalized to
a   collective normalization by the discriminant (determinant)
\begin{eq}{l}
\hbox{for }\U(n):~~
\sprod{\det W}{\det W}=\det\sprod~~=1\cr
\end{eq}

The invariance group of a scalar product in diagonal bases
\begin{eq}{l}
\U(n)=\{U\in\GL(\C^n)\mid U^\star\bl1_n U=\bl1_n\}\cr
\end{eq}is equivalent  to the invariance group of the
scalar product, reoriented by $W$ in a particle basis
 to $W^\star\bl1_nW=\sprod~~$
\begin{eq}{l}
\{G\in\GL(\C^n)\mid G^\star\sprod~~G=\sprod~~\}=W^{-1}\U(n) W
\end{eq}i.e. $W$ is determined up to $\U(n)$.

A remark with respect to the general structure which is used
at different points in established physical theories:
An orientation manifold $G/H$ arises with the distinction of a subgroup
$H$ in a  group $G$. Prominent examples are -
for spacetime operations with the distinction of a Lorentz group $\SO_0(1,3)$
- the  manifold of Lorentz metrics $g$ as factorized by
the tetrad\cite{WEYL29} $h$
\begin{eq}{l}
g_{\mu\nu}(x)=h_\mu^i(x)\eta_{ij}h_\nu^j(x),~~
h_\mu^i(x)\in\GL(\R^4)/\SO_0(1,3)
\end{eq}and - for internal operations in the standard model of elementary
particles\cite{WEIN} -
the Goldstone manifold $\U(2)/\U(1)_+$
arising with the distinction of an electromagnetic
group $\U(1)_+$ in the hyperisospin group $\U(2)$ as
illustrated by the Higgs isodoublet $\Phi$ as related
to the Goldstone transmutator $U$ (discussed in more detail in \cite{S011})
\begin{eq}{l}
\Phi^\star_\al\Phi^\be=(U^\star)_\al^a
({\bl 1_2-\tau_3\over 2})_a^bU_b^\be,~~U_b^\be\in\U(2)/\U(1)_+
\end{eq}

As well as, e.g., a spacetime translation in Minkowski space
$x\in\R^4$ has to be seen as a `whole'  where vector components make sense
only if
an additional physical structure distinguishes a coordinate system,
e.g. rest systems for massive particles,
a probability collective including decaying particles
(translation eigenstates with complex energies), e.g. the two neutral kaons,
together with their decay products, has a holistic identity.
In addition to the particle bases which have a physical relevance there exist
many orthogonal bases, e.g.  CP-eigenstates for the kaon system,
 which are not related to particles.
What can be really measured are transition amplitudes between particles states,
not between fictive other states.
This is
in analogy to the many
special relativistic spacetime decompositions into time and
space or into lightlike and spacelike subspaces,
as illustrated above with Sylvester and Witt bases, which have  sense
for measurements
only if, e.g., base determining  particles is considered.
Via the nonvanishing transition elements
the identity of the energy eigenstates is spread over the whole collective.
Obviously, for a small width ${\Ga\over m}\ll 1$, the
uncomplete identity of a decaying particle - its collective property -  will be difficult to discover.
However, it is to be expected that there  are experiments which can test
the  difference between
an individualistic probability interpretation of decaying
particles and their collective probability interpretation
where the collective discriminant normalization may be relevant.

\vskip10mm
\centerline{\bf Acknowledgement}
\vskip4mm

I am indebted to
Walter Blum for discussions leading to the concept of
 probability collectives for
unstable particles.


\begin{thebibliography}{99}

\bibitem{BRS}
{C. Becchi, A. Rouet, R. Stora, {\it Ann. of Phys.} 98 (1976), 287}

\bibitem{BOE}{H. Boerner, {\it Darstellungen von Gruppen} (1955), Springer,
      Berlin, G\"ot\-tin\-gen, Heidelberg}

\bibitem{ALG9}{N. Bourbaki, {\it Alg\`ebre, Chapitre 9}
(Formes sesquilineaires et formes quadratiques), Hermann,  Paris (1959)}

\bibitem{KO}{T. Kugo, I. Ojima, {\it Progr. of Theor. Phys.} 60 (1978), 1869}



\bibitem{WEIN}{S. Weinberg, {\it Phys. Rev. Lett.} 18 (1967), 507}

\bibitem{WEYL29}{H. Weyl, {\it Z. Physik} 56 (1929), 330}
\bibitem{WIWEI}{V.F. Weisskopf, E. Wigner, {\it Z. Physik} 63 (1930), 54}









\bibitem{S89}{H. Saller, {\it Nuovo Cimento} 104B (1989), 291}

\bibitem{S911}{H. Saller, {\it Nuovo Cimento} 104A (1991), 493;
106B (1992), 1319; 106A (1993), 469}


\bibitem{S922}{H. Saller, {\it Nuovo Cimento} 108B (1993), 603
and 109B (1993), 255}




\bibitem{S011}{H. Saller, {hep-th/0103043},
{Spacetime as the Manifold of the Internal Symmetry Orbits in the External
Symmetries}}


 \end{thebibliography}
\end{document}